\documentclass{aa}  
\usepackage{graphicx}
\usepackage{natbib}
\usepackage{amssymb}
\usepackage{amstext}
\usepackage{latexsym} 
\usepackage[sumlimits]{amsmath}
\newcommand{\Ha}{H$\alpha$}			%H-alpha
\newcommand{\NII}{[N{\sc ii}]}			%[NII]
\usepackage{txfonts}
\begin{document}
   \title{Star-forming galaxies in low-redshift clusters: Comparison
     of integrated properties of cluster and field galaxies \thanks{
Based on observations made with the Nordic Optical Telescope,
operated on the island of La Palma jointly by Denmark, Finland, Iceland,
Norway, and Sweden, in the Spanish Observatorio del Roque de los 
Muchachos of the Instituto de Astrof\'isica de Canarias; and with 
the Jacobus Kapteyn Telescope, which was operated on the island of La Palma
by the Isaac Newton Group in the Spanish Observatorio del Roque de los
Muchachos of the Instituto de Astrof\'isica de Canarias. 
}}
   \titlerunning{Comparison
     of integrated properties of cluster and field galaxies}

	\author{C.F.~Bretherton\inst{1,2}
	 \and P.A.~James\inst{1}
	 \and C.~Moss\inst{1}\thanks{Deceased 12th May 2010}
         \and M.~Whittle\inst{3}}

   \institute{Astrophysics Research Institute, 
	Liverpool John Moores University, 
	Birkenhead CH41 1LD
	\and Carter Observatory, Wellington, New Zealand
        \and Department of Astronomy, University of Virginia, 
Charlottesville, VA 22903, USA}

   \date{Received ; accepted }

% \abstract{}{}{}{}{} 
% 5 {} token are mandatory
 
  \abstract
  % context heading (optional)
  % {} leave it empty if necessary  
   {}
  % aims heading (mandatory)
   {We investigate the effect of the cluster environment on the 
star formation properties of galaxies in 8 nearby Abell clusters.}
  % methods heading (mandatory)
   {Star formation properties are determined for individual galaxies
using the equivalent width of \Ha$+$\NII\ line emission from narrow-band 
imaging.  
Equivalent width distributions are derived for each galaxy type in 
each of 3 environments - cluster, supercluster (outside the cluster
virial radius) and field.  The effects of morphological disturbance
on star formation are also investigated. }
  % results heading (mandatory)
   {We identify a population of early-type disk galaxies in the cluster
population with enhanced star formation compared to their field counterparts.
The enhanced cluster galaxies frequently show evidence of disturbance,
and the disturbed galaxies show 
marginal evidence for a higher velocity dispersion, possibly indicative of an infalling population.
}
  % conclusions heading (optional), leave it empty if necessary 
   {}


   \maketitle
%
%________________________________________________________________

\section{Introduction}

The transformation of spiral galaxies to lenticulars in the cluster
environment over the last $\sim$5 Gyr is now well established
\citep{but78, but84, dre97}, but the processes responsible for
bringing about this change are still the subject of vigorous
discussion.  One set of mechanisms centres on the removal of gas from
spiral galaxies in the cluster environment, whether through
collisional sweeping \citep{spi51, val90}, ram-pressure stripping of
disk gas \citep{gun72, cow77, nul82, qui00}, or the removal of
large-scale gas reservoirs in `strangulation' or `starvation'
scenarios \citep{lar80, bal00, bow04}, with the latter postulated to
result in anaemic spirals \citep{bek02}. Many studies have looked at
the effect of tidal interactions either in the general cluster
environment \citep{nog86, lav88, hen96}, or in cluster sub-units
\citep{her95, bar96, bek99, gne03}.  Some have concluded that tidal
effects lead to galaxy-wide star formation \citep{byr90}; this is one
of the possible consequences of repeated high-velocity galaxy-galaxy
encounters, sometimes termed `harassment' \citep{moo99, mih04}.
Low-velocity tidal encounters, on the other hand, tend to drive gas
into the central regions of galaxies leading to nuclear star formation
(SF) and the build-up of bulges \citep{ken87, mih92, ion04}.  It is
also possible that the bulk of the evolutionary activity took place
through pre-processing in galaxy groups \citep{zab98} prior to the
assembly of these groups into the present-day clusters.

It should be noted that, given the range of processes likely to be
operating, it is not clear {\it a priori} whether the predominant
effect of the cluster environment will be to suppress (through gas
removal or exhaustion) or enhance (through, e.g., tidal triggering) total SF
rates in disk galaxies.  Observational studies have adduced evidence
supporting both possibilities, with suppressed SF being found by both
\citet{bal98} and \citet{has98}, whereas others \citep{don90, mos93, gav94,
biv97, mos98, gav98} conclude that cluster spirals have SF activity similar
to or enhanced in comparison with the field population.

A comprehensive survey of the observational work undertaken in this
area is beyond the scope of this paper, but it is illustrative to
consider some of the approaches that have been adopted.  One route is
to focus on the properties of galaxies in intermediate-redshift
clusters, as exemplified by \citet{mor07} who studied the SF activity
in clusters at redshifts of $\sim$0.5. They identify a population of
disk galaxies with young stellar populations but no ongoing SF, which
they take as evidence for gradual curtailment of SF through
`strangulation'or similar processes.  However, they also identify a
density threshold for intra-cluster gas, above which a single passage
of even a large galaxy can lead to gas stripping and an abrupt
transformation to a gas-depleted lenticular with no star formation.

A second approach is to use the statistical properties of very large
numbers of low-redshift galaxies now made available by the
Sloan Digital Sky Survey (SDSS).  \citet{par07} present a study of the
colours and morphologies of $>$300,000 SDSS galaxies within
$z\sim$0.1, which they correlate with the local number density
around each of the galaxies. They find the
fraction of galaxies with early morphological type to be a
monotonically increasing function of this number density, and
propose tidal processes as the dominant mechanism of galaxy
transformation.

A third method, and the one adopted in the present study, is to look
in great detail at the SF properties (rates and spatial
distributions) of spiral galaxies in the nearest clusters.  An
excellent example of this approach is the study of the Virgo cluster
undertaken by \citet{koo04b, koo04a} using spatially resolved SF
mapping based on \Ha\ narrow-band imaging.  They find that many of the
Virgo cluster spiral galaxies show outer truncation of their SF, in
comparison with a sample of field galaxies, and some show
centrally-enhanced SF.  They conclude that a combination of
ram-pressure stripping and tidally-induced SF are required to
explain these observations.

The present paper is part of a study that is applying techniques
similar to those of \citet{koo04b, koo04a} to eight other nearby
galaxy clusters.  The sample definition is presented in the first
paper of this series \citep{tho08} which also describes the
observations (broad- and narrow-band CCD photometry) and the data
reduction process.  The present paper contains an analysis of the
total SF properties (rates and \Ha\ equivalent widths) of the cluster
galaxies, which are compared with a field galaxy
sample derived from the \Ha\ Galaxy Survey \citep{jam04}, henceforth
\Ha GS.  Future papers will look at more detailed properties of SF
within these galaxies, e.g. concentration indices and radial
distributions.

\section{Data}

\subsection{The comparison samples}

Global parameters for all observed cluster galaxies are given in Table
2 of \citet{tho08}. However, for a robust comparison with the \Ha GS
field data, the cluster data are restricted to two well defined
subsamples. The first of these is a complete sample of all Sa--Sc
galaxies in six of the eight survey clusters (Abell 400, 426, 569,
779, 1367, 1656) as surveyed by the Objective Prism Survey
\citep[henceforth OPS;][]{mos00,mos05}.  The second contains all
emission line galaxies (ELGs) detected by the OPS in all eight
clusters (i.e. the above six clusters plus Abell 262 and 347) which
excludes some of the galaxies with emission lines of lower equivalent 
width \citep{tho08}.

A full discussion of the completeness of the ELG sample as a function
of \Ha\ flux, equivalent width (EW) and surface brightness is given in 
\citet{tho08}.  To summarise, all three factors affect the detectability
of galaxies by the OPS, with surface brightness being the most important.
The ELG sample becomes significantly incomplete below an EW of 2~nm, and 
below an \Ha\ flux of 3.2$\times 10^{-17}$~W~m$^{-2}$.  However, the cleanest
detectability threshold is given by \Ha\ surface brightness with a limit
of 4$\times$10$^{-20}$~W~m$^{-2}$~arcsec$^{-2}$.

An essential requirement for this project is consistent morphological
classifications across the different samples used. For many of the
cluster galaxies, no literature classifications were available, and
these were provided by one of the authors (MW) working from plate
material as explained in \citet{mos00}. These classifications were
done on the revised de Vaucouleurs system \citep{dev59,dev74}, and
intercomparisons were performed where possible, with galaxies with
classifications given in the UGC \citep{nil73} and RC3 \citep{dev91}
catalogues.  These comparisons revealed no systematic offsets and a
scatter of about 1 $T$-type in classification, similar to the scatter
found from blind repeats of the same galaxies.  For the field sample,
classifications were again on the de Vaucouleurs system, taken
directly from the RC3 or UGC.

\citet{mos06} suggests that the late type (Sa and later) cluster galaxy
population has an infalling component with higher velocity dispersion
than the earlier types, as well as an asymmetric velocity dispersion
relative to the cluster mean. On the other hand, the early type
(E-S0/a) galaxies are consistent with a virially relaxed population
with a Gaussian velocity distribution. \citet{mos06} therefore
determines revised cluster mean velocities and velocity dispersions
from only the early type objects, using biweight estimators of scale
and central location. Assuming that the distribution of galaxies
follows the mass distribution of the cluster and that the system is
spherically symmetric, and following \citet{lew02}, Moss estimates
virial radii for all clusters using:

\begin{equation}
r_{vir} \simeq 3.5\sigma(1+z)^{-1.5},
\end{equation}

where $r_{vir}$ is in units of Mpc for $\sigma$ in units of 1000~km~s$^{-1}$. 
This gives values 40\% larger than the $R_{200}$ cluster radius often used 
as a proxy for the virial radius \citep[e.g. by][]{fin05} for standard 
cosmological parameters ($H_0=$70~km~s$^{-1}$~Mpc$^{-1}$, 
$\Omega_{\Lambda}$=0.7, $\Omega_{0}$=0.3); both definitions give a direct
proportionality between virial radius and cluster velocity dispersion.

Table 1 of \citet{mos06} shows revised values of cluster mean
velocity, $\overline{v}$ and velocity dispersion, $\sigma$, along with
virial radius, $r_{vir}$, for each sample cluster. These were used to
combine the sample galaxies, from six or eight individual clusters,
into one ensemble cluster, where radial distances from the centre are
normalised by the virial radius, and velocities relative to the
cluster mean are normalised by the cluster's velocity dispersion.

Galaxies within the two subsamples are restricted to those with
velocities within 3$\sigma$ of the revised mean cluster velocity,
$\overline{v}$. The data are also split into cluster galaxies, which
lie within 1 $r_{vir}$ of the cluster centre, and supercluster field
galaxies lying beyond 1 $r_{vir}$. The Sa--Sc sample comprises 105
galaxies of which 5 have no detectable \Ha\ emission. The ELG sample
comprises 115 objects, all of which have been detected in emission in
the current CCD data. Known AGN were excluded by searching the
NASA/IPAC Extragalactic Database (NED) and removing all galaxies
classified as Seyfert (Sy, with any numerical subtype) or LINER. Two
galaxies lying close to bright stars have also been excluded.

Galaxies in the cluster and supercluster subsamples are also
classified, based only on their $R$ band images, as to whether they show
signs of tidal disturbance. Objects exhibiting strong tidal features
and/or obvious distortion are classified as $T$, tidally disturbed;
those with less obvious warps, probable tidal tails and/or some
disturbance are given a classification of \textit{T:}, probably
disturbed; an asymmetric appearance or slight distortion of outer
spiral arms leads to a classification of \textit{T::}, possibly
disturbed; and galaxies with no sign of tidal disturbance are assigned
no value in this category.

The field data are taken from \Ha GS, a study of the SF properties of
327 field galaxies of all spiral and irregular types (S0a - Im, barred
and unbarred), with apparent magnitudes brighter than m$_B=$15.5,
recession velocities less than 3000~km~s$^{-1}$ and diameters between
1.7 and 6.0 arcminutes. The full \Ha GS sample includes a large number
of low-luminosity galaxies, but the subsample studied here was
restricted to galaxies brighter than M$_{B} \sim-$18.5.  All galaxies
classified as AGN in NED (26 in total in \Ha GS) were excluded from
the sample, as was done for the cluster sample, but it is worth noting
that a study of the mean \Ha\ emission profiles of the \Ha GS galaxies
\citep{jam09} found that central unresolved components are not common,
and typically contribute less than 10\% of the total emission-line
flux when they are present. Thus even if some low-level AGN are
included in the cluster or field samples (as will almost certainly be
the case), the effect on amounts and spatial distributions of emission
line flux should be small.  The \Ha GS survey does include some Virgo
cluster and group galaxies, which have also been excluded from the
comparison sample. The final field comparison sample includes 65
galaxies, 50 of which are of types Sa--Scd, with 4 of type S0a and the
remaining 11 being late-type spirals, Sd-Sm. Absolute $B$-band
magnitudes for the field data were taken uncorrected from \Ha GS and
have been corrected for internal and external extinction.  Internal
extinction corrections $A_B$ were calculated following the methods of
\citet{dev91}, including both inclination- and type-dependence as
follows:

$$A_B = \alpha(T)log(a/b)$$

where $(a/b)$ is the major-to-minor axis ratio, here evaluated at the 
$\mu_R=$24 isophote, and 

$$\alpha(T)=1.5-0.03(T-5)^2.$$

For galaxies with T$<$0, no internal extinction correction was applied, 
and for peculiar galaxies and those without spiral subtypes, a mean 
spiral correction of 1.3 mag was used.  Although the resulting
overall shapes of the cluster and field M$_{B}$ distributions differ, they are
well matched in mean and range.

Comparing the morphological type distributions of the cluster and
field samples, a larger fraction of early type and fewer late type
spirals are seen in the cluster environment, as expected from the
morphology-density relation. The cluster ELG sample also contains
three type categories not covered by the \Ha GS survey. Three ELGs
have types E--S0 (S0$-$), 15 are classified as peculiar (Pec), which
lie outside the Hubble sequence, and 23 are spirals of uncertain type
(S...). The ELG sample also includes two galaxies with no type
information, which are excluded from much of the following analysis.

\Ha GS observations are restricted to galaxies with major-to-minor
axis ratios of less than or equal to 4.0, in order to exclude highly
inclined objects ($i \gtrsim 81^{\circ}$), where extinction effects are
likely to be strongest. However, a Kendall rank test on the complete
cluster Sa--Sc sample shows no significant dependence ($\tau = -0.09$,
probability = 0.16) of EW on inclination within this survey, and
therefore no axis ratio cut is applied to the cluster samples.

\subsection{Observational data}

The data used are derived from CCD images taken in broad-band $R$ and
narrow-band \Ha\ filters, using the 1.0 m Jacobus Kapteyn Telescope
(JKT) and the 2.6 m Nordic Optical Telescope (NOT), both situated on
the island of La Palma.  The instrumentation used on both the JKT and
NOT and the resulting cluster galaxy photometry are described fully in
\citet{tho08}, and the equivalent for the comparison field sample in
\citet{jam04}, so the details will not be repeated here.  It should be
noted that the line fluxes and equivalent widths used are for the
\Ha\ line and the neighbouring \NII\ lines.  The narrow-band imaging
was continuum-subtracted using appropriately-scaled and aligned
$R$-band images, which generally gives good results but inevitably
gives substantial errors for galaxies with low equivalent width
emission.

\section{Distribution of equivalent widths}
\label{sec:ewd}

\begin{figure}
\includegraphics[width=87mm]{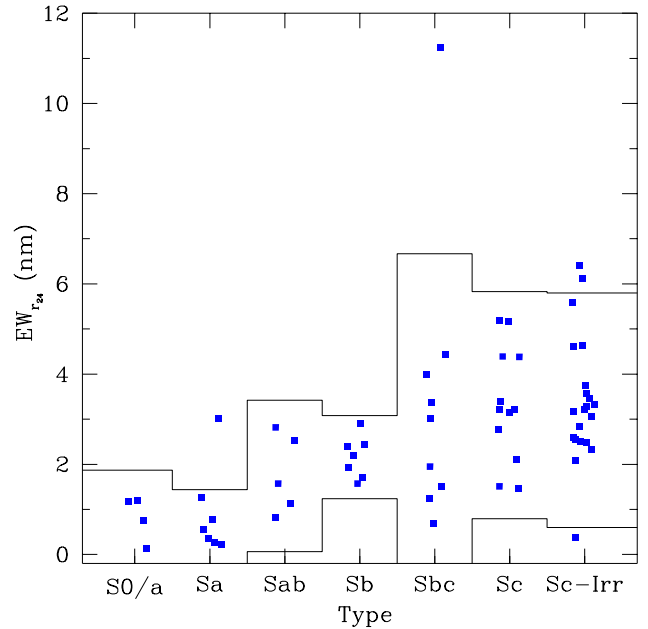}
\caption{Global EW vs. morphological type for the full field
  comparison sample. Solid lines show the 2$\sigma$ limits of the
  field population. Typical errors in EW are 10--15\% for high EW ($>
  2$ nm) and 25--35\% for low EW ($< 2$ nm) galaxies
  (\citealt{sha02}).}
\label{fig:biw_F}
\end{figure}

\begin{figure}
\includegraphics[width=87mm]{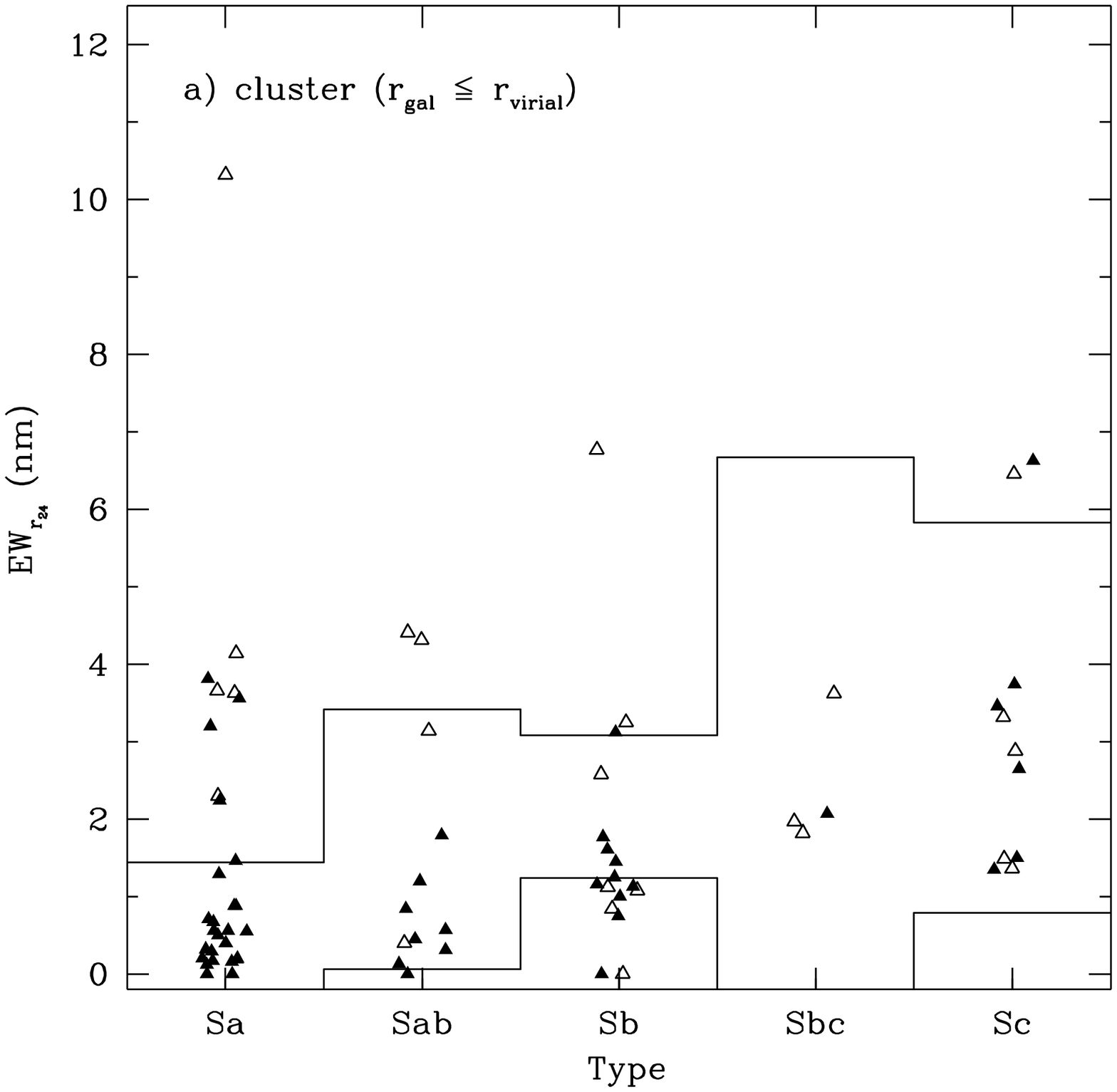}
\includegraphics[width=87mm]{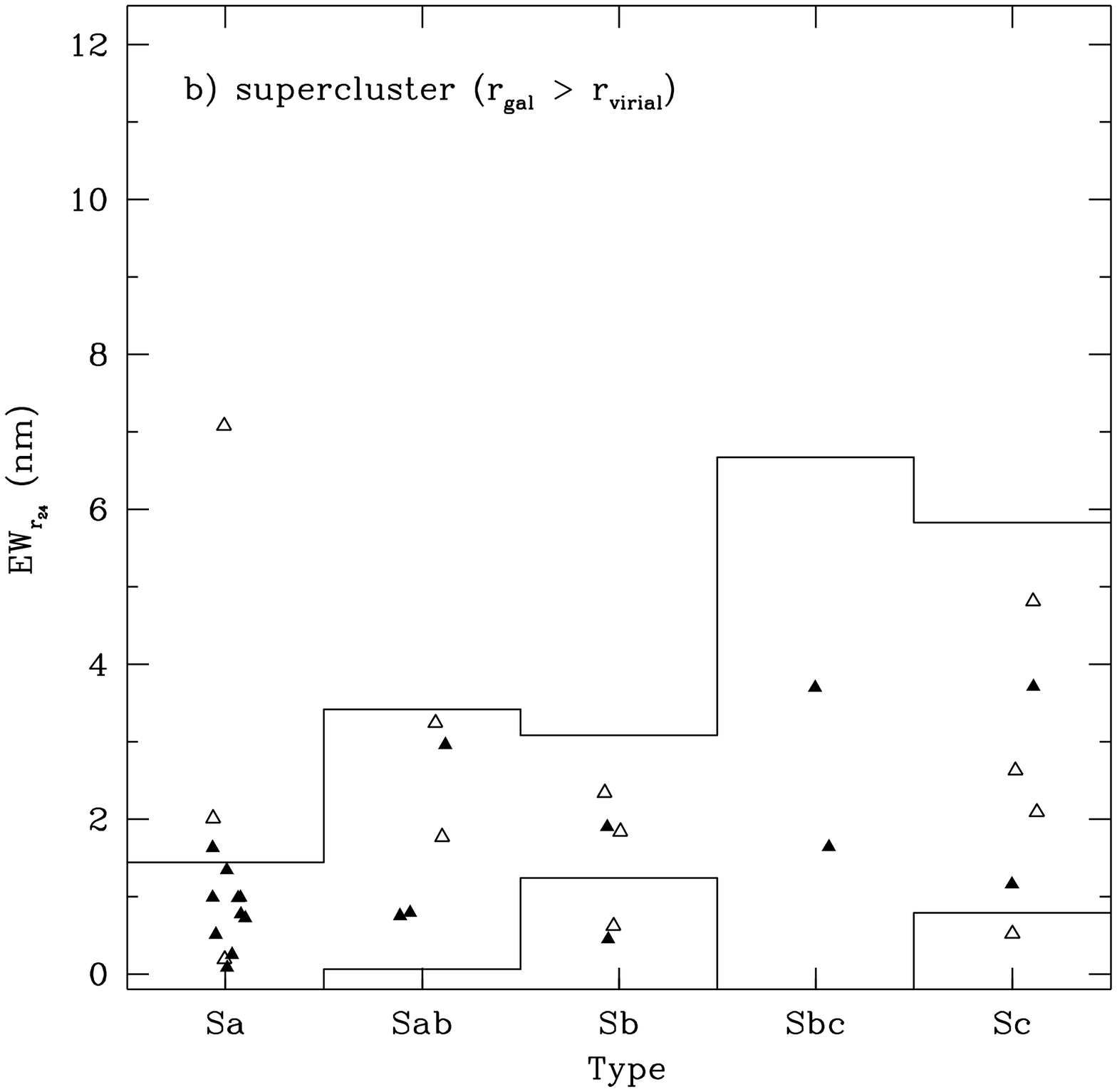}
\caption{Distribution of EW with Hubble Type for the cluster Sa--Sc sample
  split into disturbed (open triangles) and undisturbed (filled
  triangles) objects. Solid lines show the 2$\sigma$ field limits from
  Figure \ref{fig:biw_F}. Typical errors in EW are $\sim$5--15\% for
  high EW ($> 2$nm), 15--25\% for moderate EW (1--2 nm) and 25--100\%
  for low EW ($<$ 1 nm) objects.}
\label{fig:biw_Sac}
\end{figure}

Figure \ref{fig:biw_F} shows the distribution of total EW values with
type for the comparison field sample, where the total EW for each
galaxy is taken as the EW within the $R=$ 24 mag/sq. arcsec isophotal
radius ($r_{24}$). A biweight estimator method is used to calculate
the mean EW and standard deviation for each type and the 2$\sigma$
limits of the field data are plotted as solid lines in the figure.

The majority of field points lie within these limits. It is worth
noting here that the field galaxy with the highest EW in the field
sample, and which lies well beyond the 2$\sigma$ limit, is the Sbc
galaxy UGC5786 (NGC 3310). This is a well-studied example of a local
UV-bright starburst with a complex peculiar morphology
(\citealt{kin93}; \citealt{con00}), including a ``bow and arrow''
structure in the outer regions (\citealt{wal67};
\citealt{bal81,ber84,mul95}) most likely caused by a recent
merger with a smaller galaxy (\citealt{bal81,mul96};
\citealt{con00}).

Figure \ref{fig:biw_Sac} shows similar plots for the cluster and
supercluster Sa--Sc samples, where again the solid lines show the 2$\sigma$
limits generated from the field sample, and galaxies are split into
disturbed (open points) and undisturbed (filled points) objects on
each plot. It can be seen from Figure \ref{fig:biw_Sac}a that a
substantial fraction of cluster galaxies lie beyond the 2$\sigma$
field limits ($\sim$36\% compared to only 5\% of field Sa-Sc
galaxies). Ten cluster Sa--Sc galaxies (13.5\% of the total) appear to
have EW reduced compared to the field sample. Three of these galaxies
have no detected \Ha\ emission, however, 9 of the low emission objects are
Sb spirals for which the range in field EW values is surprisingly
small.

More significant are the 17 galaxies (23\%), particularly of earlier
Sa--Sb types, with EW values beyond the upper field limits. This suggests
that the cluster environment is causing an enhancement of star
formation in some spirals. There is also some evidence that galaxies
with enhanced emission may preferentially be disturbed, with nearly
59\% of enhanced galaxies showing clear signs of tidal disturbance
compared to less than 23\% of galaxies with no enhancement in EW
($\chi^{2}$ probability $< 0.05$). The supercluster field sample on
the other hand is similar to the true field population, with only two
galaxies lying significantly above the 2$\sigma$ limits, and both of these
have a disturbed appearance.

\begin{figure}
\includegraphics[width=87mm]{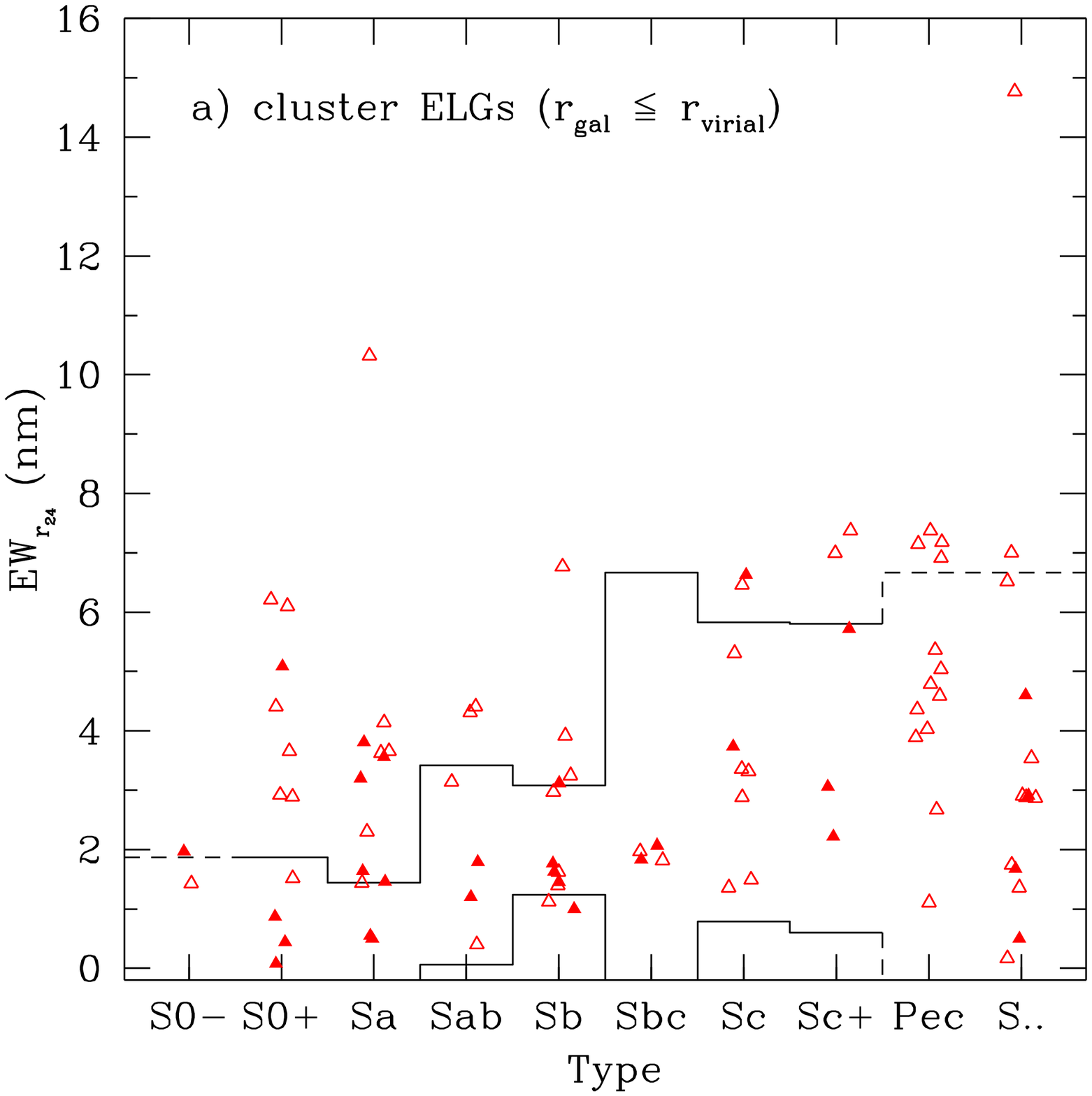}
\includegraphics[width=87mm]{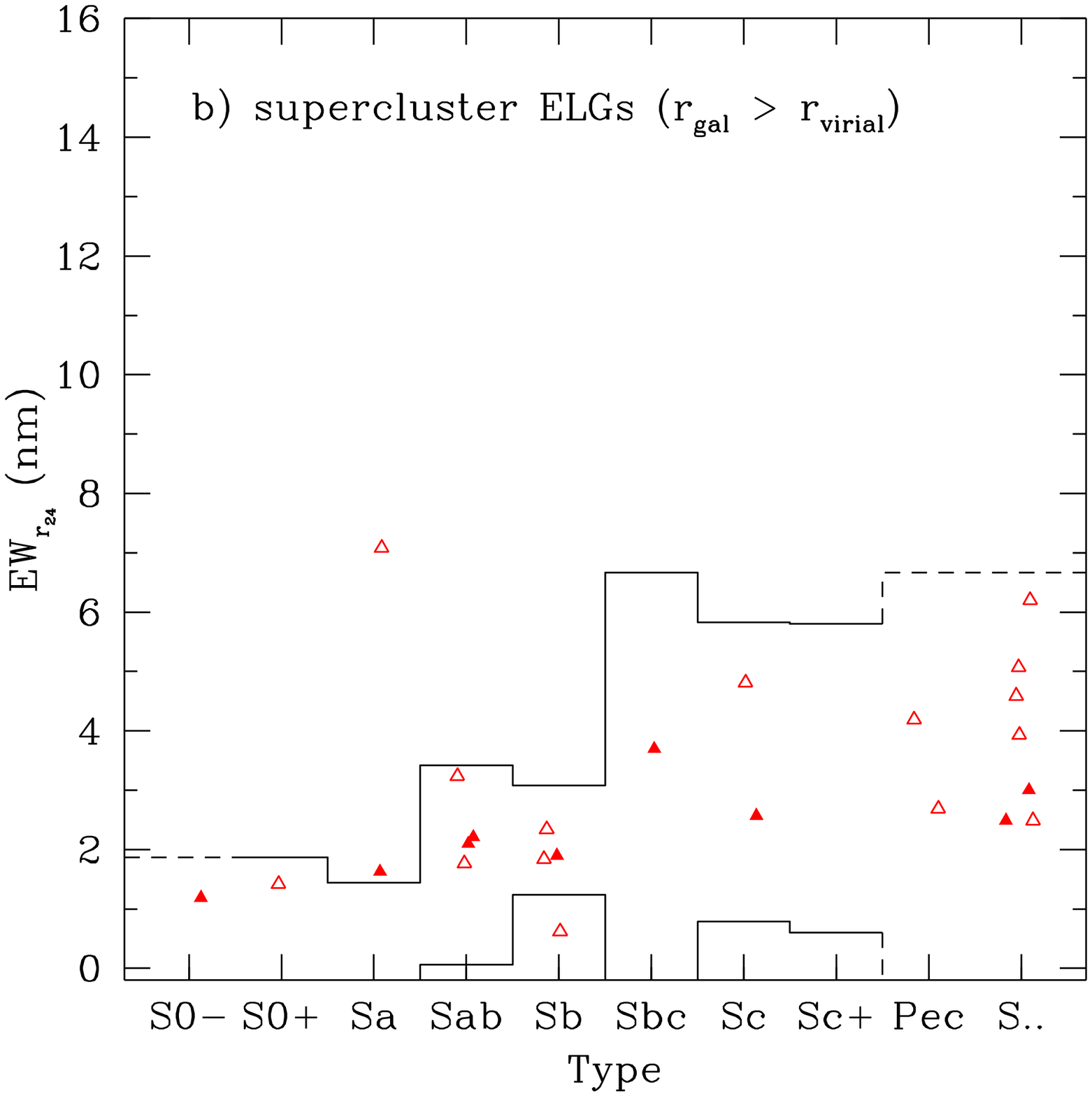}
\caption{As Figure \ref{fig:biw_Sac} but showing the distribution of
  EW with Hubble Type for the ELG sample. Dashed lines show assumed
  limits for ELG types not included in the field sample (see text for
  details). Typical EW errors are the same as Figure
  \ref{fig:biw_Sac}.}
\label{fig:biw_elg}
\end{figure}

Figure \ref{fig:biw_elg} shows the distribution of EW values with type for
the cluster ELG sample, again split into cluster and supercluster
field objects. Here the solid lines again show the 2$\sigma$ limits
for the field sample. For the peculiar galaxies and
spirals of uncertain type, the upper limit is taken as the Sbc value
(shown as a dashed line) as this is the highest 2$\sigma$ value in the
field sample and should therefore provide a conservative limit for
these categories, such that any peculiar or unknown spiral galaxies
with EW values above this point can also be considered enhanced. For the
three E--S0 (S0$^-$) galaxies in the ELG sample it is assumed that the
upper field limit is the same as that for the S0--S0/a (S0$^+$)
objects. Again this is likely to be higher than the true value as
E--S0 types generally have very little or no current star formation.

Once again, Figure \ref{fig:biw_elg}a shows an increase of EW for a
number of cluster galaxies, particularly of types Sa--Sb, and
this enhanced emission also extends to earlier types with seven of 11
S0--S0/a objects (64\%) having EW values enhanced relative to the field. It
can also be seen that a few late type Sc and later galaxies as well as
some peculiar and unclassifiable spiral galaxies appear to have enhanced
emission. In contrast, the supercluster galaxies in Figure
\ref{fig:biw_elg}b are almost indistinguishable from the true field
sample.

The ELG sample is biased towards objects with more luminous, higher
surface brightness \Ha\ emission, and so includes fewer galaxies with
low EW. The low emission objects seen in the Sa--Sc sample are,
therefore, not included in the ELG dataset. However, of the galaxies
included in the ELG sample, some 34 objects (38\%) have enhanced EW
values, of which 25 ($\sim$74\%) are classified as disturbed. The
non-enhanced ELG objects also have a relatively high proportion of
disturbed galaxies, with around 61\% showing signs of tidal
disturbance. Even excluding peculiar galaxies, which are, as expected,
all disturbed, this figure still stands at 53\%, much higher than the
23\% disturbed galaxies seen amongst the Sa--Sc non-enhanced
objects. This suggests that disturbed galaxies preferentially have
brighter \Ha\ emission, even for non-enhanced objects.

\begin{figure}
\vspace{-20mm}
\includegraphics[width=87mm]{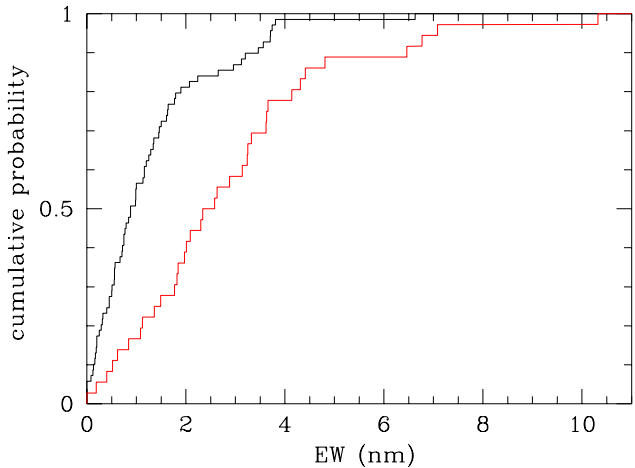}
\caption{Cumulative distributions with EW for disturbed (red) and
  undisturbed (black) galaxies. A K-S test suggests the distributions are
  significantly different.}
\label{fig:ew_cp}
\end{figure}

Assigning values 0--3 to the tidal disturbance categories from
undisturbed (no rank) to definitely disturbed ($T$) allows a Kendall
rank test to be carried out on the complete Sa--Sc sample. This shows
a substantial correlation ($\tau = 0.39$) of EW with tidal disturbance
which is significant at the $>5\sigma$ level. Similarly, a K-S test is
performed to compare the distribution of EW values in the disturbed
and undisturbed samples. This gives a probability of only
$1.1\times10^{-5}$ that the data are drawn from the same distribution,
showing that the disturbed and undisturbed objects have significantly
different EW distributions at $>4\sigma$ level. The cumulative
distributions with EW for the disturbed (red) and undisturbed (black)
Sa--Sc galaxies are shown in Figure \ref{fig:ew_cp}.

\citet{koo04a} also detect a number of spirals with star formation rates
enhanced by up to a factor of 3. They find that these are generally
lower luminosity galaxies (M$_{B} > -18$), for which they lack a good
field comparison sample. These authors therefore restrict their sample
to galaxies brighter than M$_{R_{24}} = -19.5$ (M$_{B} \sim -18.5$) to
avoid a possible luminosity bias. \citet{ken84} and \citet{bos01} also
find that lower luminosity galaxies tend to have higher EW values.

For the full Sa--Sc cluster and supercluster data, a Kendall rank test
shows a moderate ($\tau = 0.24$) but significant (3.7$\sigma$)
dependence of EW on $R$ band magnitude. This result is in agreement
with, for example, \citet{gav96}, who find a significant
anti-correlation between galaxy mass and specific star formation rate.
The current sample, however, has a limiting magnitude of M$_{B}\sim
-18.5$, equivalent to the cut made by \citet{koo04a}, and therefore
includes only relatively bright galaxies. The mean magnitude of the
cluster sample is also slightly brighter than that of the field
galaxies. The correlations found in the present study, and by
\citet{gav96}, between EW and galaxy luminosity would thus predict
overall {\em lower} EW values for the cluster sample, and cannot be
used to explain the enhanced emission of cluster Sa--Sc galaxies when
compared to the field.

\section{Mean EW values}
\begin{figure}
\includegraphics[width=87mm]{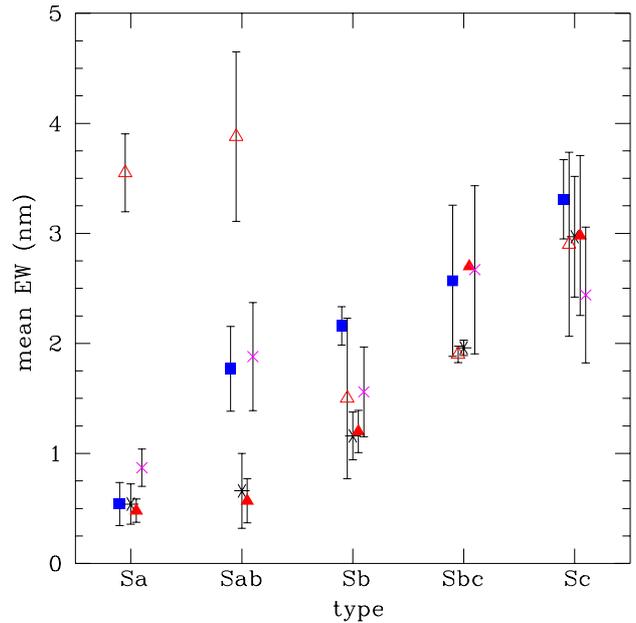}
\caption{Mean EW values with type for the Sa--Sc cluster sample (stars), the
  field (squares), and the supercluster field (crosses). The cluster
  sample is also split into disturbed (open triangles) and undisturbed
  (closed triangles) galaxies.}
\label{fig:sacmean}
\end{figure}

Figure \ref{fig:sacmean} shows the mean EW values with type for the Sa--Sc
cluster (stars), supercluster (crosses) and field (squares)
samples. Means are calculated using biweight estimators to reduce the
influence of outliers and the error bars shown give the standard error
on the mean, $\sigma/\sqrt{n}$.  The cluster data are also split into
disturbed (open triangles) and undisturbed (closed triangles)
galaxies. This split is not done for the supercluster sample due to the small
numbers of objects involved (only 31 galaxies in total), however, the
mean EW values for the full supercluster sample are generally consistent
with the field data.

It can be seen from Figure \ref{fig:sacmean} that the mean EW values for the
full cluster data seem to be reduced in general compared to the field
sample, with the exception of Sa galaxies, but this is only
significant for Sab and Sb types. The undisturbed objects follow the
same trend, and with the exception of Sbc types, where only a single
object is undisturbed, the undisturbed mean EW values are very similar to
those for the full sample. The disturbed objects, on the other hand,
show a very different trend, with mean EW values for Sa--Sab galaxies
greatly enhanced above both total cluster and field values. For later
types, however, the mean disturbed EW values are close to, and consistent
with, both the full and undisturbed cluster samples. Although the
highest EW enhancements are seen for the disturbed early type spirals,
the fraction of Sa--Sab galaxies that are tidally disturbed is only
21\%, compared to 48\% and 45\% for Sb--Sbc and Sc galaxies
respectively.

\begin{figure}
\includegraphics[width=87mm]{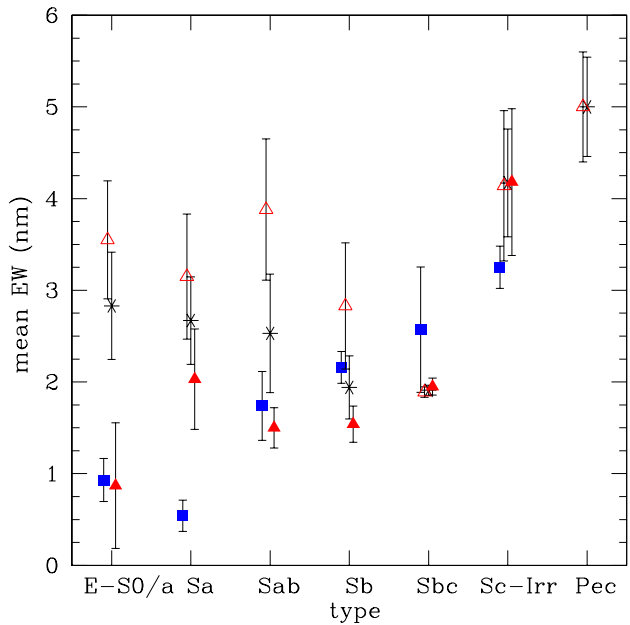}
\caption{Mean EW values with type for the cluster ELG sample (stars)
  split into
  disturbed (open triangles) and undisturbed (filled triangles)
  galaxies. The field data are shown as filled squares. Points for the
  supercluster subset are not shown, because of the small number
  of galaxies in this category; see Fig.~\ref{fig:biw_elg}. }
\label{fig:elgcmean}
\end{figure}

A similar plot is shown in Figure \ref{fig:elgcmean} for the cluster
ELG sample. Here all E--S0/a cluster galaxies are grouped into a
single bin and compared to the field mean S0/a EW. In order to
increase the numbers of objects in the cluster later type bins, all
Sc--Irr galaxies have also been grouped together. Spirals of
unclassifiable type are not included in this figure. The undisturbed
cluster ELGs have EW values comparable to the field data, however, the
ELG sample is biased towards galaxies with brighter \Ha\ emission, so
the true mean EW values may be somewhat lower. Once again, the
disturbed galaxy sample shows significantly enhanced mean EW values
for Sa--Sab spirals, but this also extends further to even earlier
types (E--S0/a). The sample of peculiar galaxies, all of which are
disturbed, has a mean EW greater than any field type. The increases of
mean EW in specific Hubble types in Figs. \ref{fig:sacmean} and
\ref{fig:elgcmean} illustrate that the difference in the overall EW
distributions of the disturbed and undisturbed galaxies, shown in
Fig. \ref{fig:ew_cp}, cannot be attributed simply to differences in
the morphological makeup of the disturbed and undisturbed samples.
Disturbance seems to affect EW in galaxies of a given type, at least
for early types.

It is of interest to consider possible reasons for this enhancement in EW
being apparent only for early types.  Two explanations can be proposed.
The first is that enhanced emission tends to take place in the densest 
cluster regions, where we expect galaxies to be stripped and generally have
lower disk emission, thus appearing as early-type galaxies.  The second is 
that the scatter in EW values for late-type unstripped galaxies is much 
larger than for early types, and may therefore tend to mask any effect.

\section{Clustercentric radial distribution of \Ha\ EW}
\label{sec:ccrd}

\citet{koo04a} conclude that truncation of the star forming disk, via
ICM--ISM stripping, is the dominant process affecting galaxies in
clusters. If this is correct then the mean EW of cluster spiral
galaxies should decrease towards the cluster centre.

\begin{figure}
\includegraphics[width=87mm]{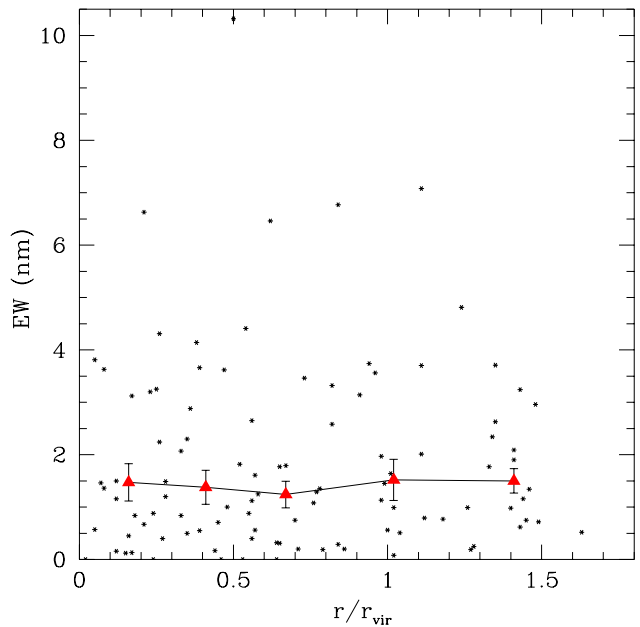}
\caption{\Ha\ EW as a function of clustercentric distance for the
  Sa--Sc sample. Large points show mean distances and EW values for bins of
  equal numbers of galaxies.}
\label{fig:vrad}
\end{figure}

Studies using the 2dF Galaxy Redshift Survey (\citealt{lew02}) and
SDSS (\citealt{gom03}) have suggested that there is a transition in
star formation activity at a characteristic density corresponding to
the local density at $\sim$1 virial radius, although it is difficult
to disentangle this from the known morphology--density relation
(\citealt{bos06}). \citet{gav06} also find evidence that the average
\Ha\ EW of luminous spirals in the Virgo and Coma + Abell 1367 cluster
samples decreases in the inner $\sim$1 virial radius, although the
binning of the data results in very few points within 1~$r_{vir}$ (two
points for Virgo, one for Coma + A1367) such that it is not possible
to trace any gradual variation within the cluster
itself. \citet{yua05}, however, studied the star formation properties
of 184 bright cluster galaxies in the $z\sim0.08$ cluster A2255 and
found that, although there is a slight trend for the specific star
formation rates of early-type galaxies to decrease towards the cluster
centre, the inner late-type galaxies, in fact, tend to have higher
star formation rates.

Figure \ref{fig:vrad} shows the EW values of the Sa--Sc sample plotted
against distance in virial radii from the composite cluster
centre. Individual galaxies are plotted as black stars, and the red
triangles show the mean distance and equivalent width in 5
approximately equal bins, calculated using a biweight estimator. The
error bars show the standard error in EW for each bin. The mean EW values in
the cluster and supercluster Sa--Sc sample are lower than the overall
field mean for the Sa--Sc sample, however, this is likely to be due to
the different morphological mix in the cluster and field
samples. Figure \ref{fig:vrad}, however, shows no change in the star
formation rate of late type galaxies with clustercentric radius. This
suggests that stripping alone cannot dominate the transformation of
spiral galaxies in these clusters.

It is tempting to suggest that the lack of an observed trend between
decreasing star formation and proximity to the cluster centre may be
due to field interlopers projected towards the inner parts of the
cluster. \citet{mos00}, however, find that contamination by field
galaxies accounts for only $\sim$20\% of spirals within 0.5 Abell
radii (which corresponds to roughly 0.5 $r_{vir}$ for the composite
cluster), but will be more important outside this radius. Field
contamination would therefore likely increase any such trend and
cannot account for the flat distribution observed.

\begin{figure}
\includegraphics[width=87mm]{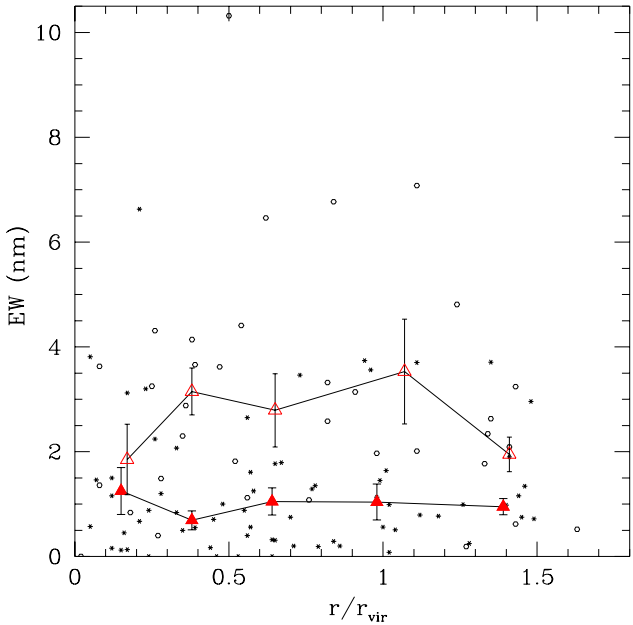}
\caption{As Figure \ref{fig:vrad}, but split into disturbed (open
  points) and undisturbed (closed points) galaxies.}
\label{fig:vradd}
\end{figure}

\begin{figure}
\includegraphics[width=87mm]{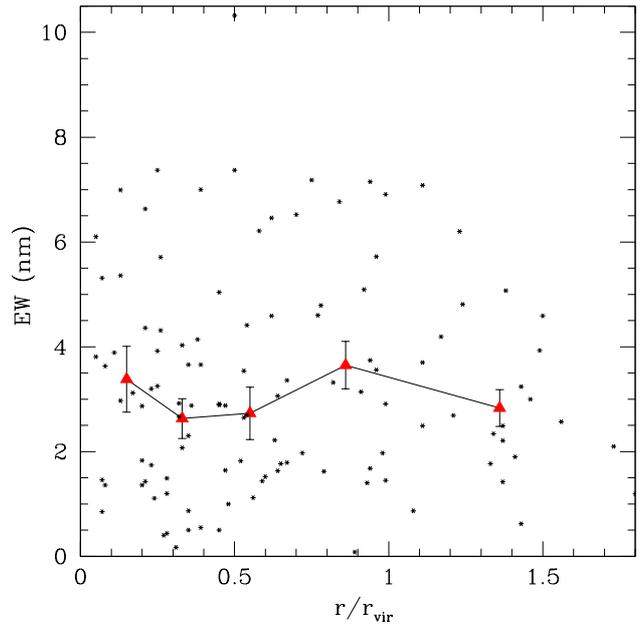}
\caption{As Figure \ref{fig:vrad}, but for the ELG sample.}
\label{fig:vradelg}
\end{figure}

Given the observed differences in star formation activity for
disturbed and undisturbed galaxies, it is also of interest to study
the clustercentric radial distributions of EW values for each sample
separately. This is shown in Figure \ref{fig:vradd}. Here individual
undisturbed galaxies are shown as before with black stars, whilst the
disturbed population have open points. As with previous plots, the
mean values for the undisturbed galaxies are plotted as filled
triangles whilst open triangles represent the means for the binned
disturbed objects. Figure \ref{fig:vradd} shows that, as with the
complete sample, no change is seen in mean EW with distance from the
cluster centre for the undisturbed sample. As expected, the means for
the disturbed galaxies are higher than those for the undisturbed
objects, but, although the mean star formation appears to be slightly
higher at intermediate clustercentric distances, the points and error
bars are still consistent with a flat distribution. The same lack
  of a significant trend is found for the ELG sample, shown in Fig.
  \ref{fig:vradelg}.

This result is initially surprising, given the number of studies
finding negative correlations between emission-line strength and the
number density of the local environment.  This dates back at least as
far as \citet{ost60}, who found emission lines to be less prevalent in
elliptical galaxies in dense clusters than in Virgo cluster
ellipticals.  The same trend was found for larger samples of cluster
galaxies, including spiral galaxies, by \citet{gis78} and
\citet{dre85}.  Most recently, \citet{vul10} conclude that the average
SF rate in 604 galaxies within 16 intermediate-redshift clusters vary
systematically with environment, even at fixed galaxy mass.

However, other studies have found different results, more in line
with those found here. These include \citet{biv97}, who found that
their overall conclusions regarding the correlation between
emission-line strength and environment depended critically on a
systematic bias resulting from the different effects of
magnitude-limited selection on field and cluster samples.  Once this
had been corrected for, they found no difference between the emission
line properties of field and cluster galaxies of a given morphological
type.  This result was confirmed by \citet{mos05}, who further discuss
the selection effect analysis of \citet{biv97} and find results
similar to those of the present paper for 379 galaxies in low-redshift
galaxies, with Objective Prism measures of \Ha\ emission.  Finally,
\citet{car01} and \citet{rin05} both conclude that the distributions
of total \Ha\ EW values for galaxies with significant SF show no
difference between samples selected within or outside the virial
radius of their host clusters.  

\section{Distribution of disturbed cluster galaxies}

A comparison of the cumulative distributions of the clustercentric distances of
disturbed vs. undisturbed Sa--Sc galaxies suggests that the radial
distributions of these samples are similar. A K-S test gives a
probability of 0.89 that they are drawn from the same parent
distribution.

%\begin{figure} \vspace{-20mm}
%\includegraphics[width=84mm]{rvir_cp.eps} 
%\caption{Cumulative distribution functions with clustercentric
%distance for disturbed (red) and undisturbed (black) Sa--Sc
%galaxies.}
%\label{fig:rvir_cp} \end{figure}

\begin{figure}
\includegraphics[width=128mm]{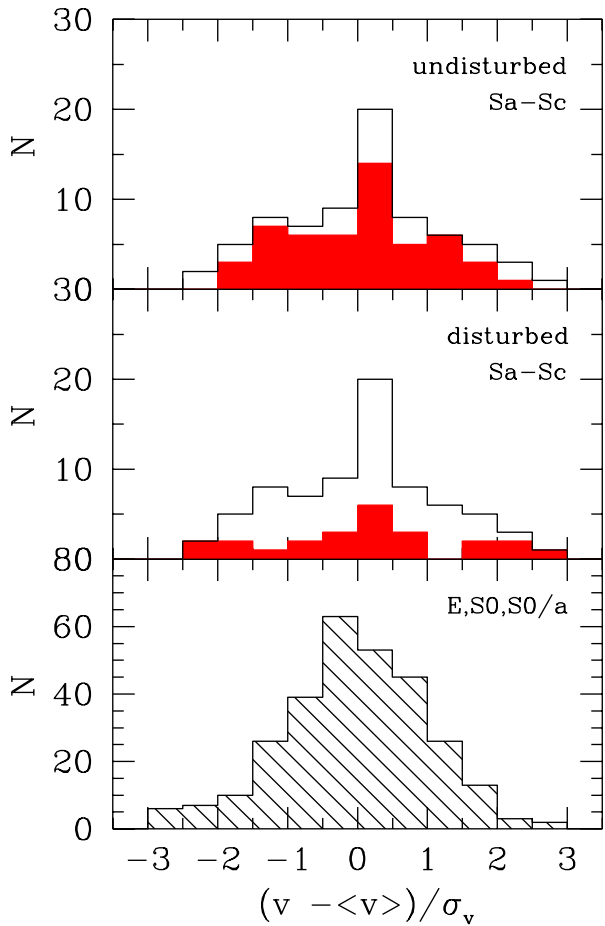}
\caption{Normalised velocity distribution for Sa--Sc sample galaxies 
(top and centre), scaled to the cluster mean. Open histograms show the 
full cluster Sa--Sc sample, with filled histograms showing the disturbed 
(centre) and undisturbed (top) populations. The bottom plot shows the 
normalised velocity distribution for the early type population in the six 
clusters included in the Sa--Sc sample.}
\label{fig:vhd}
\end{figure}

Figure \ref{fig:vhd}, however, shows a rather different conclusion
based on the distribution of galaxy velocities within their
clusters. The normalised velocity dispersion is shown for all Sa--Sc
sample galaxies within 1~$r_{vir}$ of the composite cluster centre
(open histogram). The sample is also split into disturbed and
undisturbed galaxies, and their distributions are plotted separately
in the filled red histograms in the centre and top panels
respectively. The undisturbed galaxies appear to be fairly centrally
peaked, while the disturbed galaxies show a much flatter
distribution. The bottom panel in Figure \ref{fig:vhd} shows the
normalised velocity distribution for the early type elliptical and
lenticular galaxies in the six Sa--Sc sample clusters. These objects
were used to calculate the cluster means and dispersions that were
then employed to normalise the later type population, and they
therefore have a mean normalised velocity of 0.0, with a standard
deviation of 1.0.

The mean and dispersion of the velocity distribution have been
calculated for each Sa--Sc sample using biweight estimators. The
undisturbed sample has a mean of $0.11 \pm 0.14$ with a dispersion of
1.02, and is therefore consistent with the early type population used
to calculate the cluster velocity distributions (K-S probability =
0.77, data folded about $(v-\langle v \rangle )/\sigma_{v} = 0$). For
the disturbed galaxies, however, although the mean value of $0.09 \pm
0.30$ is still consistent with the composite cluster mean, the
dispersion has a much higher value of 1.45. This is, in fact, likely
to be a lower limit to the dispersion of disturbed galaxies as the
$3\sigma$ velocity cut applied in the selection of the sample means
that galaxies with higher velocity deviations would have been
omitted. Even so, a K-S test shows that the velocity distributions of
the disturbed Sa--Sc sample and the early type population are
different at $\gtrsim2\sigma$ significance. The distribution of
undisturbed galaxies drops off before the $3\sigma$ limit. Comparing
the velocities of the disturbed and undisturbed Sa--Sc galaxies, a K-S
test gives a probability of 0.06 that these are drawn from the same
distribution. The marginally higher velocity dispersion observed for
the disturbed galaxies ($\sim\sqrt{2}$ greater than the undisturbed
sample) is suggestive of an infalling population, and a similar result
was found for the larger sample of cluster galaxies studied by
\citet{mos06}. It should also be noted that the disturbed galaxy
Sa--Sc sample studied here has a somewhat later mean type ($T=$2.96,
cf. 2.12 for the undisturbed galaxies) which may have some bearing on
the interpretation of this result.

\section{Summary}
\label{sec:summary}

Comparison of the global \Ha\ EW values of cluster, supercluster and
field galaxies has identified a population of cluster galaxies
(particularly early type spirals and lenticulars) with enhanced star
formation compared to their field counterparts. These objects are also
more likely to have a disturbed appearance than non-enhanced galaxies.
Tidal disturbance is found to be correlated with higher \Ha\ EW at
$>5\sigma$ (Kendall rank test). A K-S test also shows that the
distributions of EW values in disturbed and undisturbed populations
are significantly different ($>4\sigma$).  This disturbance is seen in
the stellar component and hence is indicative of tidal effects, rather
than, e.g., ram-pressure stripping.  A number of galaxies with
unusually weak line emission are also seen in the cluster Sa--Sc
sample. The supercluster samples, on the other hand, appear very
similar to the field.

Comparing the mean EW values for each type suggests that star
formation may be reduced in general for undisturbed cluster galaxies
across most types. The disturbed galaxies, however, have mean EW
values well above those for field early type spirals, but for later
types they are consistent with the field and undisturbed cluster
samples. The sample of highly disturbed, peculiar galaxies included in
the ELG sample has a mean EW higher than any other field or cluster
type. These results suggest that galaxy--galaxy interactions and
mergers may play a significant role in the evolution of cluster
spirals.

A study of the clustercentric radial distribution of \Ha\ EW also
shows no correlation between EW and distance from the cluster
centre. This suggests that stripping alone, which would lead to a
gradual decrease of star formation towards the cluster centre, cannot
be solely responsible for the transformation of spiral galaxies in
these clusters, and adds further weight to the argument that tidal
interactions between galaxies may also be important.

An investigation of the distribution of disturbed and undisturbed
galaxies within the cluster shows that, although there appears to be
no difference in distribution as a function of clustercentric
distance, the disturbed galaxies have a marginally higher velocity
dispersion that may indicate an infalling population.

Future papers in this series will look in more detail at the distribution
of SF activity within the galaxies studied in the present paper.  This
analysis will initially use concentration indices as a measure of the 
compactness of SF, to probe the prevalence of outer truncation and 
centrally-concentrated starbursts.  Radial light profiles will then be 
used in a more detailed study of these processes.
%__________________________________________________________________
\begin{acknowledgements}
This research
has made use of the NASA/IPAC Extragalactic Database (NED) which is
operated by the Jet Propulsion Laboratory, California Institute of
Technology, under contract with the National Aeronautics and Space
Administration. The referee is thanked for many helpful suggestions.
CFB, PAJ and MW dedicate this paper to the memory of our respected 
and sadly-missed colleague, Chris Moss.
\end{acknowledgements}

%__________________________________________________________________
%\begin{thebibliography}{}
\bibliographystyle{bibtex/aa}
\bibliography{refs}
      
%\end{thebibliography}
%__________________________________________________________________
\end{document}